\documentclass[twocolumn,showpacs,superscriptaddress,preprintnumbers,prd]{revtex4}
\usepackage{graphicx,amsmath,dcolumn,bm}
\begin{document}
\preprint{KEK-TH-1363}
\title{Clustering aspects in nuclear structure functions}

\author{M. Hirai}
\affiliation{Department of Physics, Faculty of Science and Technology,
             Tokyo University of Science, \\ 
             2641, Yamazaki, Noda, Chiba, 278-8510, Japan}
\author{S. Kumano}
\affiliation{KEK Theory Center,
             Institute of Particle and Nuclear Studies,
             High Energy Accelerator Research Organization (KEK) \\
             and Department of Particle and Nuclear Studies,
             Graduate University for Advanced Studies, \\
             1-1, Ooho, Tsukuba, Ibaraki, 305-0801, Japan}
\author{K. Saito}
\affiliation{Department of Physics, Faculty of Science and Technology,
             Tokyo University of Science, \\ 
             2641, Yamazaki, Noda, Chiba, 278-8510, Japan}
\author{T. Watanabe}
\affiliation{Department of Physics, Faculty of Science and Technology,
             Tokyo University of Science, \\ 
             2641, Yamazaki, Noda, Chiba, 278-8510, Japan}   
\date{December 28, 2010}
\begin{abstract}
For understanding an anomalous nuclear effect experimentally
observed for the beryllium-9 nucleus at the Thomas Jefferson National
Accelerator Facility (JLab), clustering aspects are studied in 
structure functions of deep inelastic lepton-nucleus scattering 
by using momentum distributions calculated in antisymmetrized 
(or fermionic) molecular dynamics (AMD) and also in a simple shell model
for comparison. According to the AMD, the $^9$Be nucleus consists
of two $\alpha$-like clusters with a surrounding neutron.
The clustering produces high-momentum components in nuclear 
wave functions, which affects nuclear modifications of 
the structure functions.
We investigated whether clustering features could appear 
in the structure function $F_2$ of $^9$Be along with studies
for other light nuclei.
We found that nuclear modifications of $F_2$ are similar in both 
AMD and shell models within our simple convolution description
although there are slight differences
in $^9$Be. It indicates that the anomalous $^9$Be result
should be explained by a different mechanism from the nuclear binding 
and Fermi motion. If nuclear-modification slopes $d(F_2^A/F_2^D)/dx$ are 
shown by the maximum local densities, the $^9$Be anomaly can 
be explained by the AMD picture, namely by the clustering structure,
whereas it certainly cannot be described in the simple shell model. 
This fact suggests that the large nuclear modification in $^9$Be
should be explained by large densities in the clusters.
For example, internal nucleon structure could be modified
in the high-density clusters.
The clustering aspect of nuclear structure functions is 
an unexplored topic which is interesting for future investigations.
\end{abstract}
\maketitle

\section{Introduction}\label{intro}

Nuclear modifications of structure functions $F_2$ were found
by the European Muon Collaboration (EMC) \cite{emc03}, so that
the phenomena is often called the EMC effect. 
Such modifications are now measured from relatively small $x$
($\sim 10^{-3}$) to large $x$ ($x \sim 0.8$),
where $x$ is the Bjorken scaling variable.
By using the data on nuclear structure functions, 
optimum parton distribution functions (PDFs) are proposed 
for nuclei \cite{npdfs,recent-npdfs}.
Physics mechanisms are different depending on the $x$ region
for producing the nuclear modifications.
At small $x$, suppression of $F_2$ occurs and it is known
as nuclear shadowing. It is due to multiple scattering of
a $q\bar q$ pair coming from the virtual photon. 
At medium and large $x$, modifications are understood 
by conventional models mainly with nuclear binding and Fermi motion
of nucleons. However, it may not be possible to explain full 
experimental modifications by such mechanisms, which indicates 
that internal structure of the nucleon could be also modified 
in a nuclear medium. For explanations of these physics mechanisms, 
the reader may look at Ref. \cite{sumemc}. 

In future, much details of the nuclear modifications will be
investigated in lepton-nucleus deep inelastic scattering
and hadron-hadron reactions.
For example, a nuclear modification difference between
up and down valence quark distributions ($u_v$ and $d_v$)
will be investigated by measuring cross sections
of semi-inclusive $\pi^\pm$ productions \cite{uvdv-exp}.
It could lead to a possible solution of the long-standing NuTeV
weak-mixing angle ($\sin\theta_W$) anomaly \cite{nutev-anomaly}
from a viewpoint of the nuclear modification difference
between $u_v$ and $d_v$ \cite{sinth}.
In addition, nuclear shell structure of the EMC effect, so called 
``local EMC effect" \cite{local-emc}, will be investigated 
by measuring semi-inclusive reactions \cite{local-exp}.
There is also an issue of the nuclear-modification 
difference between the structure functions of charged-lepton
and neutrino reactions \cite{neutrino-A,mstw08}. It needs to 
be solved for a precise determination of nucleonic and
nuclear PDFs. Such nuclear effects will be investigated
by neutrino reactions of the MINER$\nu$A project \cite{minerva}.
There will be also measurements at hadron facilities at 
RHIC (Relativistic Heavy Ion Collider),
Fermilab (E906 experiment) \cite{e906},
LHC (Large Hadron Collider),
and possibly at J-PARC (Japan Proton Accelerator Research Complex)
\cite{j-parc}.

Measurements on the EMC effect at the Thomas Jefferson
National Accelerator Facility (JLab) obtained an anomalous result
for the beryllium-9 nucleus in comparison with measurements 
for other light nuclei \cite{jlab-2009}. It is anomalous 
in the sense that the magnitude of the nuclear-modification slope 
$| \, d(F_2^A/F_2^D)/dx \, |$ is much larger in $^9$Be than
the ones expected from its average nuclear density. 
From the experimental $^9$Be radius, namely the average
nuclear density, modifications of $F_2$ at medium $x$ are
expected to be much smaller than the ones of $^{12}$C,
whereas measured values are similar in magnitude.

It is known in nuclear structure studies that
the $^9$Be nucleus has a typical cluster-like structure rather
than a shell-like one \cite{amd-1995,amd-1997}. 
It is like a cluster of two $\alpha$ ($^4$He nucleus)
particles with surrounding neutron clouds according to the studies
of antisymmetrized molecular dynamics (AMD).
This fact indicates that there exist higher-density regions
than the ones expected from the average density by the shell model
or the one estimated by the experimental charge radius. 
The high-density regions could contribute to larger 
nuclear modifications of the structure function $F_2$.
It could be a reason for the anomalous modification for $^9$Be.
Such a cluster structure could produce high-momentum
components in the momentum distribution of the nucleon,
which is eventually reflected in modifications of quark momentum
distributions, namely the structure functions of nuclei.

These considerations motivated us to investigate cluster aspects
in the structure functions $F_2^A$ for light nuclei, especially $^9$Be. 
At this stage, there is no theoretical work on the nuclear-clustering
aspect in high-energy nuclear processes, for example, in structure 
functions, although there are some studies on multi-quark clusters such as
a six-quark state in 1980's. Our current studies are totally different
from these works on multi-quark effects. 
In this article, we investigate possible nuclear clustering effects
on the structure functions $F_2^A$ within a convolution model
for describing $F_2^A$ by using the AMD and shell-model wave functions.

In. Sec. \ref{formalism}, our theoretical formalism is provided for
describing nuclear structure functions $F_2^A$. First, the convolution
model is introduced. Then, the AMD description and a simple shell model
are explained for calculating nuclear wave functions.
In Sec. \ref{results}, calculated nuclear densities are shown
for $^4$He and $^9$Be in order to illustrate the clustering structure
in the $^9$Be nucleus. Then, momentum distributions are shown for
these nuclei. The ratios $F_2^A/F_2^D$ are calculated in both AMD
and shell models, and they are compared with experimental data.
Finally, the nuclear modification slopes $d(F_2^A/F_2^D)/dx$ are
discussed. Our results are summarized in Sec. \ref{summary}.

\section{Formalism}\label{formalism}

We explain a basic formalism for calculating the nuclear structure functions
$F_2^A$ in the convolution approach together with 
antisymmetrized molecular dynamics and a simple shell model
for calculating nuclear wave functions.
These models are somewhat obvious within each community in 
structure-function and nuclear-cluster physicists.
However, the following introductory explanations are intended that
different communities could understand with each other.

\subsection{Nuclear structure functions in convolution approach}
\label{convolution}

The cross section of deep inelastic charged-lepton-nucleon (or nucleus)
scattering is expressed by a lepton tensor $L^{\mu\nu}$
multiplied by a hadron tensor $W_{\mu\nu}$: 
$d \sigma \sim L^{\mu\nu} W_{\mu\nu}$ 
\cite{sumemc,sf-introduction,roberts,sk-pr}.
The hadron tensor is defined by 
\begin{equation}
W_{\mu\nu} (p, q) = \frac{1}{4 \pi}
       \int d^4 \xi \, e^{iq \cdot \xi} 
       < p \, | \, [J_\mu (\xi), J_\nu (0)] \, | \, p > ,
\label{eqn:define-w}
\end{equation}
where $q$ is the virtual photon four-momentum, 
$p$ is the momentum of the nucleon (or nucleus), 
and $J_\mu$ is the hadronic electromagnetic current.
The hadron tensor $W_{\mu\nu}$ is expressed by 
the imaginary part of the forward virtual Compton amplitude 
$T_{\mu\nu}$ as $W_{\mu\nu}=Im(T_{\mu\nu})/(2 \pi)$ by the optical
theorem.

\begin{figure}[b]
\includegraphics[width=0.33\textwidth]{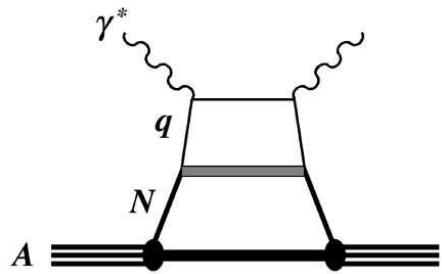}
\vspace{-0.2cm}
\caption{Convolution approach for nuclear structure functions.
The $\gamma^*$, $q$, $N$, and $A$ indicate the virtual $\gamma$,
quark, nucleon, and nucleus, respectively. A quark momentum
distribution is described by the integral of
a corresponding quark distribution convoluted with 
a nucleon momentum distribution.}
\label{fig:convolution}
\end{figure}

The convolution model has been discussed in various articles
within binding models for calculating nuclear structure functions,
so that the detailed formalism should be found, for example,
in Refs. \cite{sumemc,roberts}.
It indicates that a nuclear structure function is given
by an integral of the nucleonic one convoluted with a momentum
distribution of a nucleon in a nucleus as illustrated
in Fig. \ref{fig:convolution}. 
It is written in the hadron-tensor form as
\begin{equation}
W_{\mu\nu}^A (p_A, q) = \int d^4 p_N \, S(p_N) \, W_{\mu\nu}^N (p_N, q) ,
\label{eqn:w-convolution-w}
\end{equation}
where $p_N$ and $p_A$ are momenta for the nucleon and nucleus,
respectively, and $S(p_N)$ is the spectral function
which is the energy-momentum distribution of nucleons in the nucleus.
The structure functions are generally expressed in terms of two variables
$Q^2$ and $x$ defined by
\begin{equation}
Q^2 = - q^2, \ \ \ x=\frac{Q^2}{2 M_N \nu} ,
\end{equation}
where $M_N$ is the nucleon mass, $\nu$ is the energy transfer $\nu=q^0$
in the rest frame of a target nucleus, and $q^2$ is given by 
$q^2=(q^0)^2 - \vec q^{\, \, 2}$.
In the convolution picture of Eq. (\ref{eqn:w-convolution-w}),
the process is described by two steps as illustrated in 
Fig. \ref{fig:convolution}. First, a nucleon is distributed
in a nucleus according to the spectral function $S(p_N)$ with the nucleon
momentum $p_N$, and then a quark is distributed with the momentum 
fraction $x$ in the nucleon. The overall quark momentum distribution
is given by the convolution integral of these two distributions.

The hadron tensor $W^{A}_{\mu\nu}$ is expressed in terms of 
two structure functions $W_1^A$ and $W_2^A$ as
\begin{equation}
\! 
W^{A}_{\mu\nu} (p_A, q)  =  - W^{A}_1 (p_A, q) \tilde g_{\mu\nu}
    + W^{A}_2 (p_A, q) \, \frac{\tilde p_{A \mu} 
  \, \tilde p_{A \nu}}{p_A^2} ,
\label{eqn:hadron}
\end{equation}
where $\tilde g_{\mu\nu}$ and $\tilde p_{\mu}$ are defined by
$\tilde g_{\mu\nu} =  g_{\mu\nu} - q_\mu q_\nu /q^2$ and
$\tilde p_{\mu} = p_{\mu} -(p \cdot q) \, q_\mu /q^2$
so as to satisfy the current conservation.
The structure function $F_2^A$ is related to $W_2^A$ by
$F_2^A=W_2^A p_A \cdot q/M_A$, and its projection operator is 
given by \cite{ek03,kk08}
\begin{equation}
\widehat P_2^{\, \mu\nu}  = 
   - \frac{M_A \, p_A \cdot q}{2\, \tilde p_A^{\, 2}} 
  \left ( g^{\mu\nu} - \frac{3 \, \tilde p_A^{\, \mu} \, 
                 \tilde p_A^{\, \nu}}{\tilde p_A^{\, 2}} 
  \right ) ,
\label{eqn:p2}
\end{equation}
which satisfies $\widehat P_2^{\, \mu\nu} W^A_{\mu\nu} = F_2^A$.
The mass of the nucleus is denoted by $M_A$. 
Applying the projection operator on both side of 
Eq. (\ref{eqn:w-convolution-w}), we obtain 
\cite{li-liu-brown-88,sumemc,ek03}
\begin{equation}
F_{2}^A (x, Q^2) = \int_x^A dy \, f(y) \, F_{2}^N (x/y, Q^2) ,
\label{eqn:w-convolution}
\end{equation}
where $F_2^A$ and $F_2^N$ are structure functions for the nucleus
and nucleon, and $y$ is the momentum fraction 
\begin{equation}
y   =    \frac{M_A \, p_N \cdot q}{M_N \, p_A \cdot q} 
  \simeq \frac{A \, p_N^+}{p_A^+} ,
\end{equation}
where $p^+$ is a light-cone momentum
[$\, p^+ \equiv (p^0 +p^3)/\sqrt{2} \,$].
It should be noted that the upper bound of the variables $x$ and
$y$ is $A$ for nuclei.
The function $f(y)$ indicates a light-cone momentum distribution
for the nucleon, and it is given by 
\begin{equation}
f(y) \equiv  \frac{1}{A} \sum_{i} \int d^3 p_N
     \, y \, \delta \left( y - \frac{p_N \cdot q}{M_N \nu} \right) 
    n_i \, | \phi_i (\vec p_N) |^2 ,
\label{eqn:f(y)}
\end{equation}
where $n_i$ is the number of the nucleon in the quantum state $i$,
and the summation is taken over the occupied states.
Here, the spectral function is given by
\begin{equation}
S (p_N)= \frac{1}{A} \sum_i
      n_i \, | \phi_i (\vec p_N) |^2 
     \delta \left ( p_N^0-M_A + \sqrt{M_{A-i}^{\ 2} 
                    + \vec p_N^{\ 2}} \, \right ) ,
\label{eqn:sp}
\end{equation}
where $M_{A-i}$ is the mass of residual one-hole state,
and $\phi_i (\vec p_N)$ is the wave function of the nucleon.
Here, $\vec p_N^{\ 2}/Q^2$-type higher-twist effects \cite{ek03}
are not included in the convolution equation.
The function $f(y)$ is normalized so as to satisfy
the baryon-number conservation $\int_0^A dy f(y)=1$ 
by taking $\int dp^4_N y S(p_N) = 1$
\cite{li-liu-brown-88,cfks-2007}.

The wave functions of the nucleon are calculated non-relativistically,
and then they are used for the relativistic description in obtaining
light-cone distributions by Eq. (\ref{eqn:f(y)}). It could lead to
an issue of normalizing the non-relativistic wave function because 
there is no solid relativistic framework to use the non-relativistic
functions. Here, the wave functions are normalized to satisfy
the condition $\int dy f(y)=1$, where there is an extra factor of 
$p_N^0/M_N$ in front of $|\phi (\vec p_N)|^2$.
As noticed in the third article of Ref. \cite{kulagin}, this factor 
does not appear if a mass factor ($M_N/p_N^0$) is included 
in the convolution formalism. However, such an overall normalization 
difference does not affect our results in Sec. \ref{results}.

The separation energy $\varepsilon_i$ is defined by
\begin{equation}
\varepsilon_i = (M_{A-i}+M_N) - M_A .
\end{equation}
It is the energy required to remove a nucleon from the state $i$.
In our actual calculation, we average over all the nucleons 
for estimating the average separation energy
($\varepsilon_i$$\rightarrow$$<\varepsilon>$).
If a non-relativistic approximation is applied for the expression
$\sqrt{M_{A-i}^{\ 2}+ \vec p_N^{\ 2}}$, $p_N^0$ and $<\varepsilon>$ 
are related by considering the $\delta$ function for the energy 
conservation as
\begin{equation}
p_N^0= M_N - <\varepsilon> - \frac{\vec p_N^{\, 2}}{2 M_{A-1}},
\end{equation}
where $M_{A-i}$ is replaced by $M_{A-1}$
(the ground-state mass of the $A-1$ nucleus).
It should be noted that the residual nucleus $A-i$ could be
in an excited state and that many-body breakup processes could
be also possible in the final state. Therefore, the separation
energy is, in general, not a simple difference between 
the two nuclear binding energies in the initial and final states,
since the final nucleus would not be in the ground state. 
It means that theoretical separation energies depend how 
they are estimated. For example, they 
vary depending whether models include short-range correlations 
\cite{cfks-2007,cl-1990} and many-body breakup processes 
\cite{mt-1991}. In our work, experimental separation energies 
are taken from $(e,e'p)$ and $(p,2p)$ experiments.

Equation (\ref{eqn:w-convolution}) indicates that the nuclear
structure function $F_2^A$ is split into two parts:
the light-cone momentum distribution of the nucleon
and the nucleonic structure function $F_2^N$. 
If there is no nuclear medium effect on the nucleonic
structure function $F_2^N(x,Q^2)$, nuclear modifications
should come solely from the nucleonic distribution part, which
contains the information on nuclear binding and Fermi
motion of nucleons. These effects are reflected 
in the light-cone momentum distribution of Eq. (\ref{eqn:f(y)}), 
namely in the momentum distribution of the nucleon 
and the energy-conserving $\delta$ function. 
For calculating the distribution $f(y)$, we need a realistic
model for the wave function $\phi (\vec p_N)$.
In our work, we calculate it in two theoretical models: an antisymmetrized 
molecular dynamics and a simple shell model. They are introduced
in Secs. \ref{amd} and \ref{shell}.

\subsection{Antisymmetrized molecular dynamics}\label{amd}

This work is intended to investigate a possible clustering
effect on the structure functions of deep inelastic scattering (DIS).
There is a theoretical method, antisymmetrized molecular dynamics
(AMD) \cite{amd-intro} or fermionic molecular dynamics (FMD) 
\cite{fmd-intro}, which is developed for describing clustering
aspects of nuclei as well as shell-like structure 
{\it on an equal footing}. Hereafter, we use the nomenclature
AMD for this theoretical method.

There are nuclei which exhibit density distributions of 
separate clusters. For example, the $^8$Be nucleus has
two separate peaks, which correspond to two
$\alpha$ nuclei, in its density distribution according
to a Monte Carlo calculation for the eight-body system
by using realistic $NN$ ($N$: nucleon) potentials \cite{mc-8be-2000}. 
It suggests that some nuclei tend to form $\alpha$-like clusters 
within their structure since the $\alpha$ is a tightly bound nucleus.

A simple and yet very useful and consistent theoretical method
is provided by the AMD method. The AMD has a number of advantages,
for example, that there is no assumption on nuclear structure, namely 
shell or cluster like configuration, and that simple and systematic
studies are possible from light to medium-size nuclei.
A nuclear wave function is given by the Slater determinant
of single-particle wave packets:
\begin{align}
\! \! 
\left | \Phi (\vec r_1, \vec r_2, \cdot\cdot\cdot, \vec r_A ) \right >
   & = \frac{1}{\sqrt{A!}}
            \text{det} [ \varphi_1 (\vec r_1), \varphi_2 (\vec r_2), 
                                 \cdot\cdot\cdot, \varphi_A (\vec r_A) ] 
\nonumber \\
   & \! \! \! \! \! \! \! \! \! \! \! \! \! \! \! \! 
      =  \frac{1}{\sqrt{A!}}
  \begin{vmatrix}
        \varphi_1 (\vec r_1)     &  \varphi_1 (\vec r_2)  
      & \cdot\cdot\cdot          &  \varphi_1 (\vec r_A)  \\
        \varphi_2 (\vec r_1)     &  \varphi_2 (\vec r_2)  
      & \cdot\cdot\cdot          &  \varphi_2 (\vec r_A)  \\
        \vdots                   &  \vdots  
      & \cdot\cdot\cdot          &  \vdots             \\
        \varphi_A (\vec r_1)     &  \varphi_A (\vec r_2)  
      & \cdot\cdot\cdot          &  \varphi_A (\vec r_A)  \\
  \end{vmatrix} .
\end{align}
Here, a nucleon is described by the single-particle wave function
\begin{equation}
\varphi_i (\vec r_j) = \phi_i (\vec r_j) \, \chi_i \, \tau_i,
\end{equation}
where $\chi_i$ and $\tau_i$ indicate spin and isospin states,
respectively. The function $\phi_i (\vec r_j)$ is 
the space part of the wave function, and it is assumed to be
given by the Gaussian functional form:
\begin{equation}
\phi_i (\vec r_j) = \left ( \frac{2 \nu}{\pi} \right )^{3/4}
        \exp \left [ - \nu 
           \left ( \vec r_j -\frac{\vec Z_i}{\sqrt{\nu}} \right ) ^2
             \right ] ,
\end{equation}
where $\nu$ is a parameter to express the extent of the wave packet.
The center of the wave packet is given by $\vec Z_i/\sqrt{\nu}$.
We should note that $\vec Z_i$  is a complex variational parameter.
Its real and imaginary parts indicate nucleon position and 
momentum, respectively \cite{amd-intro}:
\begin{equation}
\! 
\frac{<\phi_i| \, \hat {\vec r} \, |\phi_i>}{<\phi_i | \phi_i>}
                   =\frac{{\rm Re} \vec Z_i}{\sqrt{\nu}}, \ \ \ 
\frac{<\phi_i| \, \hat {\vec p} \, |\phi_i>}{<\phi_i | \phi_i>}
                   = 2 \sqrt{\nu} \, {\rm Im} \vec Z_i . 
\end{equation}
A nuclear state is an eigenstate of the parity, so that 
the following parity-projected wave function is used:
\begin{align}
  \left | \Phi ^\pm (\vec r_1, \vec r_2, 
                    \cdot\cdot\cdot, \vec r_A ) \right >
  = &  \frac{1}{\sqrt{2}} \big [ \,
       \left | \Phi (\vec r_1, \vec r_2, 
                   \cdot\cdot\cdot, \vec r_A ) \right >
\nonumber \\
& \pm \left | \Phi (-\vec r_1, -\vec r_2, 
                   \cdot\cdot\cdot, -\vec r_A ) \right >
                   \, \big ] .
\end{align}

As for the $NN$ interactions, we use the following potentials:
\begin{align}
\text{2-body:} \ \ 
     & V_2 = (1-m -m P_\sigma P_\tau) 
\nonumber \\       
 & \ \ \ \ \ \ \ \ 
   \times \left [  v_{21} e^{- (r/r_{21})^2} 
                 + v_{22} e^{- (r/r_{22})^2}  \right ] ,
\nonumber \\
\text{3-body:} \ \ 
     & V_3 =  v_3 \, \delta^3 (\vec r_1 - \vec r_2 ) 
                  \, \delta^3 (\vec r_2 - \vec r_3 ) ,
\nonumber \\
\text{LS:} \ \ 
     & V_{LS} = v_{LS} \left [ e^{- (r/r_{LS1})^2} 
                  - e^{- (r/r_{LS2})^2} \right ]
\nonumber \\
 & \ \ \ \ \ \ \ \ 
   \times P(^3 O)   \vec L \cdot \vec S ,
\label{eqn:amd-interactions}
\end{align}
where $m$, $v_{21}$, $v_{22}$, $r_{21}$, $r_{22}$, 
$v_3$, $v_{LS}$, $r_{LS1}$, and $r_{LS2}$
are constants. The two-body interaction part $m P_\sigma P_\tau$
indicates the Majorana term with spin and isospin exchange
operators ($P_\sigma$, $P_\tau$). The three-body part
is a contact interaction form, 
and $P(^3O)$ is the projection operator of the triplet-odd ($^3 O$)
state (spin $S$=1, angular momentum $L$=odd) in the two-nucleon 
system \cite{3odd}. The Coulomb interaction is also considered
in our analysis. The constants $m$, $v_{21}$, $\cdot \cdot \cdot$
are taken from Ref. \cite{amd-1995} except for $v_3$ and $V_{LS}$, 
which are fixed so as to reproduce binding energies
of considered nuclei under the radius constraint:
$v_{LS}=2000$ MeV, $v_3=4000$, 3300, 2000 MeV for $^4$He, $^9$Be, $^{12}$C, 
respectively. Here, we should be careful to take into account the effect of
center-of-mass motion \cite{ono-1992}.

The AMD wave functions contain the parameters $\vec Z_i$ and $\nu$,
which are determined by minimizing the system energy with
a frictional-cooling method.
Time development of {$\vec Z_i$} is described by
the time-dependent variational principle:
\begin{equation}
\delta \int_{t_1}^{t_2} dt \,
\frac{ \left < \, \Phi (Z) \, \right | \, i \frac{d}{dt} - H \,
       \left | \, \Phi (Z) \, \right > }
     { \left < \, \Phi (Z) \, \right | 
       \left .    \Phi (Z) \, \right > } =0 .
\end{equation}
It leads to the equation of motion. 
Introducing two arbitrary parameters $\lambda$ and $\mu$
for practically solving the equation of motion, we obtain
\begin{equation}
i \frac{d}{dt} Z_i = (\lambda+i\mu) 
              \frac{\partial H}{\partial Z_i^*} .
\end{equation}
Here, $\mu$ is a friction parameter which should be a negative number. 
By solving this equation, the parameters {$\vec Z_i$} are obtained.
From the obtained parameters, the densities in coordinate and
momentum spaces are calculated by
\begin{align}
\! 
\rho (\vec r) & = \left ( \frac{2 \nu}{\pi} \right )^{\! \frac{3}{2}} \!
   \sum_{i,j} \exp \left [ - 2 
   \left ( \sqrt{\nu} \vec r -\frac{\vec Z_i^* 
                 + \vec Z_j}{2} \right )^2
             \right ] 
\nonumber \\ 
& \ \ \ \ \ \ \ \ \ 
\times B_{ij} B_{ji}^{-1}, 
\label{eqn:rho-r-amd}
\\
\rho (\vec p) & = \left ( \frac{1}{2\pi\nu} \right )^{\! \frac{3}{2}} \!
   \sum_{i,j} \exp \left [ - \frac{1}{2} 
   \left \{ \frac{\vec p}{\sqrt{\nu}} - i (\vec Z_i^* 
                 - \vec Z_j) \right \}^2
             \right ]
\nonumber \\ 
& \ \ \ \ \ \ \ \ \ 
\times B_{ij} B_{ji}^{-1}, 
\label{eqn:rho-p-amd}
\end{align}
where $B_{ij} \equiv \int d\vec r \,
         \varphi_i^\dagger (\vec r) \varphi_j (\vec r)$.
This momentum distribution is used for calculating the light-cone
momentum distribution in Eq. (\ref{eqn:f(y)}).
Then, using the convolution equation of (\ref{eqn:w-convolution}),
we obtain the nuclear structure functions, which include
clustering effects described by the AMD.

\subsection{Simple shell model}\label{shell}

In order to compare with the AMD results at this stage, we also calculate 
the nuclear spectral function by using a simple shell model,
because the current wave functions and $NN$ interactions in AMD
are simple Gaussian forms. 
If much detailed studies become necessary in future,
we may consider to use more sophisticated models,
for example, a density-dependent Hartree-Fock \cite{rs-book}
or a detailed shell model such as NuShell (OXBASH) \cite{nushell}.

As a shell model, we take a simple harmonic oscillator model.
Nucleons are assumed to move in an average central potential
created by interactions of all the nucleons in a nucleus.
Then, the nucleons are treated independently with each other.
A simple and yet realistic choice of the potential is the
harmonic-oscillator type ($M_N \omega^2 r^2/2$).
Its wave function is separated into radial- and angular-dependent parts:
\begin{equation}
\psi_{n \ell m} (r, \theta, \phi) = R_{n \ell} (r) Y_{\ell m} (\theta, \phi) ,
\end{equation}
where $r$, $\theta$, and $\phi$ are spherical coordinates,
and $n$, $\ell$, and $m$ are radial, azimuthal, and magnetic quantum numbers,
respectively. 
The function $Y_{\ell m} (\theta, \phi)$ is the spherical harmonics, and
the radial wave function is given by \cite{fw-book}
\begin{equation}
\! 
R_{n \ell} (r) = \sqrt{\frac{2 \kappa^{2\ell+3}(n-1)!}
                            {[\Gamma(n+\ell+1/2)]^3}}
                  r^\ell e^{-\frac{1}{2}\kappa^2 r^2} 
                  L_{n-1}^{\ell+1/2} (\kappa^2 r^2) ,
\end{equation}
where $L_{n-1}^{\ell+1/2} (x)$ is the Laguerre polynomial, and $\kappa$ 
is defined by $\kappa \equiv \sqrt{M_N \omega}$.

\begin{figure}[t]
\includegraphics[width=0.45\textwidth]{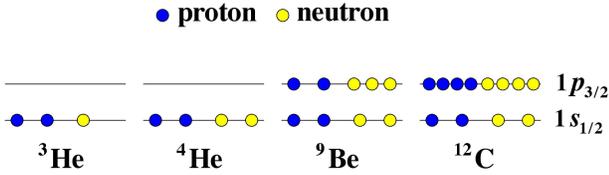}
\vspace{-0.0cm}
\caption{(Color online) Shell levels for $^3$He, $^4$He,
                       $^9$Be, and $^{12}$C.}
\label{fig:shell-level}
\end{figure}

In the following analysis, the light nuclei, $^3$He, $^4$He,
$^9$Be, and $^{12}$C are considered, so that the low-energy levels,
$1s_{1/2}$ and $1p_{3/2}$, are taken into account as shown
in Fig. \ref{fig:shell-level}.
The only parameter in the model is $\omega$, which is fixed 
by a nuclear radius. The constants of the AMD model are determined
so as to explain experimental nuclear charge radii. Then, nuclear 
matter radii are calculated by using obtained AMD densities. 
Since it is the purpose of this work to investigate a difference
between the structure functions of the AMD and shell models, 
we take the constant $\omega$ for each nucleus
so as to obtain the same matter radius calculated by the AMD.

\subsection{Deuteron wave function}\label{deuteron}

Experimental data are listed by ratios $F_2^A(x)/F_2^D$, where $F_2^D$
is the structure function of the deuteron, for showing nuclear 
modifications in the structure functions. Since the deuteron is
a bound two-nucleon system, a common wave function is used
in Eq. (\ref{eqn:f(y)}) for calculating the structure function 
$F_2^D (x,Q^2)$ of Eq. (\ref{eqn:w-convolution}) in both AMD- 
and shell-model analyses. Here, we take the deuteron wave function
given by the Bonn group in Ref. \cite{deuteron}.

\subsection{Experimental information on separation energies, binding energies,
            and charge radii}\label{data}

\begin{table}[b]
\caption{Experimental data for mean separation energies, 
binding energies per nucleon, and charge root-mean-square radii.
The matter radii are calculated in the AMD model except 
for the deuteron so that charge radii agree with the data.
The asterisk $*$ indicates a theoretical estimate of Ref.
\cite{us-1988} because the experimental data is not available.}
\label{table:energies-radii}
\centering
\begin{tabular}{@{\hspace{0.2cm}}c|@{\hspace{0.4cm}}c@{\hspace{0.4cm}}
c@{\hspace{0.4cm}}c@{\hspace{0.4cm}}c@{\hspace{0.2cm}}}
\hline
Nucleus   & $<\varepsilon>$ & $B/A$ & $\sqrt{<r^2>_c}$ & $\sqrt{<r^2>_m}$ \\
          &      (MeV)      & (MeV) &      (fm)        &     (fm)         \\
\hline
D         &  2.22           &  1.11 &      2.10        &  2.10            \\
$^3$He    &  (11.4$^*$)     &  2.57 &      1.96        &  1.96            \\
$^4$He    &  20.4           &  7.07 &      1.68        &  1.68            \\
$^9$Be    &  24.4           &  6.46 &      2.52        &  2.61            \\
$^{12}$C  &  22.6           &  7.68 &      2.47        &  2.48            \\
\hline
\end{tabular}
\end{table}

In calculating the structure functions, experimental information is 
needed for separation energies, binding energies, and charge radii.
The binding energies are taken from Ref. \cite{audia-2003} and they are
listed in Table \ref{table:energies-radii}. They are used for calculating
nuclear mass: $M_A=Z\, M_p + N \, M_n - B$, where $M_p$ and $M_n$ are
proton and neutron masses, $Z$ and $N$ are atomic and neutron numbers,
and $B$ is the binding energy.

Experimental nuclear charge r.m.s. (root-mean-square) radii are listed
for the deuteron \cite{rms-d}, $^3$He \cite{rms-3he}, 
$^4$He \cite{rms-4he}, $^9$Be \cite{rms-9be}, 
and $^{12}$C \cite{rms-12c} in Table \ref{table:energies-radii}.
Using these charge radii and binding energies, the constants 
in the AMD model are determined.
The matter r.m.s. radii are then calculated in the AMD
by using Eq. (\ref{eqn:rho-r-amd}). 
There are slight differences between $\sqrt{<r^2>_c}$ and
$\sqrt{<r^2>_m}$ in the AMD for $^9$Be and $^{12}$C.
This is due to the effect of Coulomb force. 

The separation energies are taken from experimental measurements
for $^4$He \cite{tyren-1966}, $^9$Be and $^{12}$C 
\cite{separation}, and they are listed 
in Table \ref{table:energies-radii}. 
A theoretical estimate 11.4 MeV is listed just for information 
because there is no available data for $^3$He.
It was obtained by using a spectral function calculated 
by the Faddeev method with the Reid soft-core potential \cite{us-1988}. 
 
It should be also noted \cite{3he4he} that
the separation energy 20.4 MeV of $^4$He \cite{tyren-1966}
was obtained by using the data only in the peak region 
of the energy spectrum of $^4$He(p,2p)$^3$H and 
a continuum region is not included. 
The separation energy should be calculated by
the average energy weighted by the spectral function: 
\begin{equation}
<\varepsilon> = \int dE_N d^3 p_N \, E_N S (E_N, p_N) .
\end{equation}
We notice that theoretical estimates are usually larger than 
this value (20.4 MeV) for $^4$He \cite{cl-1990,mt-1991}. 
For example, 28.2 MeV is obtained in Ref. \cite{cl-1990}, where 
the average kinetic energy estimated by the ATMS
(Amalgamation of Two-body correlation functions 
into Multiple Scattering process) method is employed,
and then the Koltun sum rule is used for estimating
the separation energy $<\varepsilon>$:
$ B/A = \left [ <\varepsilon> - <T> (A-2)/(A-1) \right ] /2 $,
where $B/A$ is the binding energy per nucleon
and $<T>$ is the average kinetic energy.
However, it is very difficult to calculate a reliable value of 
the separation energy. 
The experimental separation energies $<\varepsilon>$ were obtained
in nucleon-knockout reactions by observing peaks of single-particle 
excitations and they do not include the contribution from continuum states 
of the residual nucleus. Therefore, the mean separation energies would be
underestimated. In this work, we estimated the clustering effect without
the continuum, which needs to be considered in future for detailed
comparison with data.

\section{Results}\label{results}

\begin{figure}[b]
        \includegraphics*[width=0.40\textwidth]{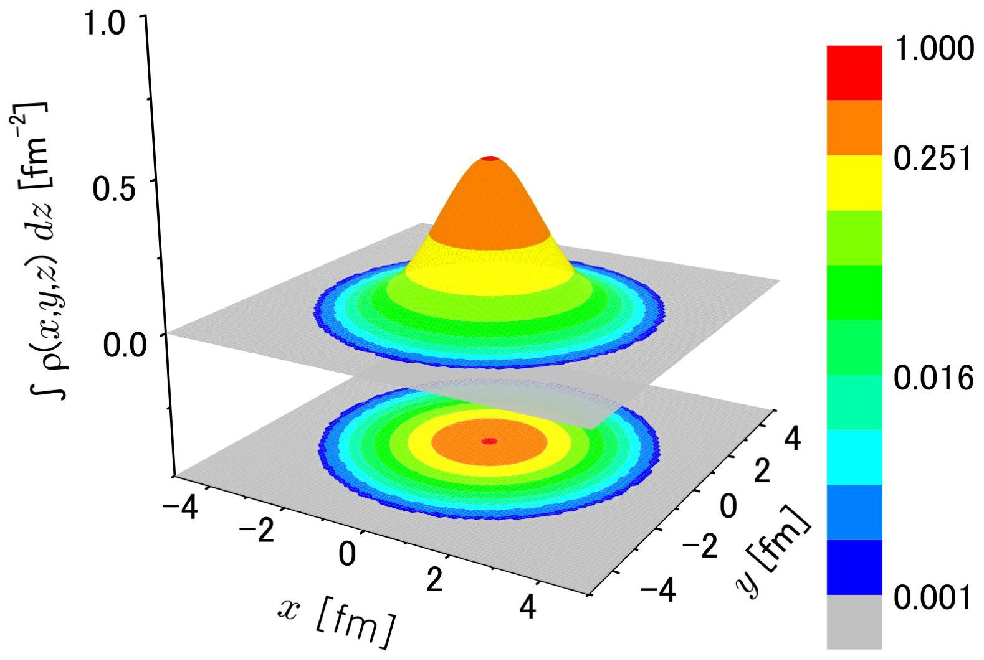} \\
        \includegraphics*[width=0.40\textwidth]{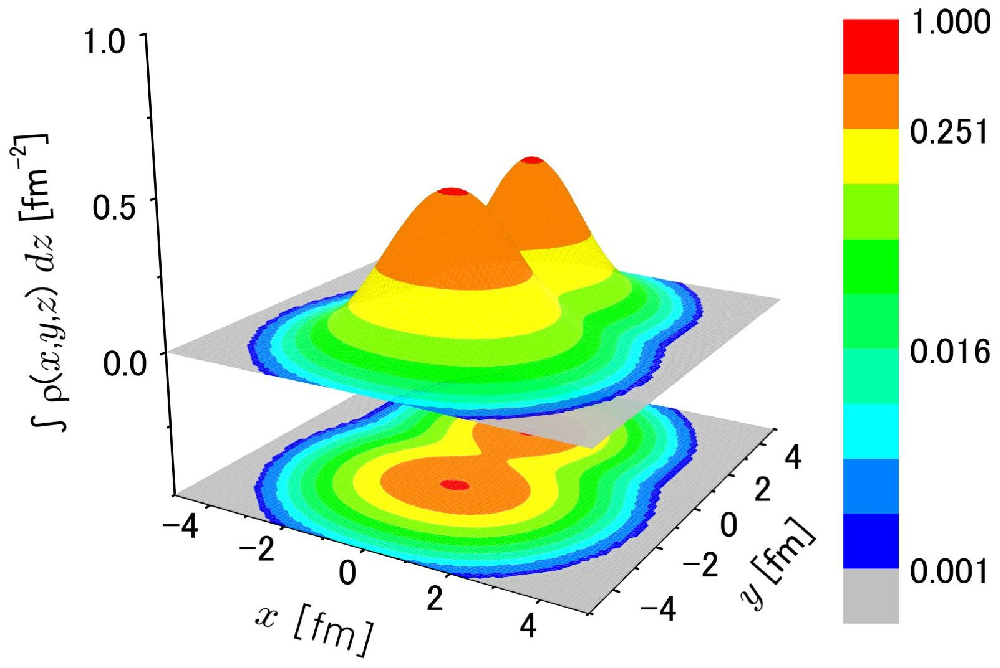} \\
        \vspace{-0.25cm}
\caption{\label{fig:amd-4he-9be}
(Color online) Upper and lower figures indicate coordinate-space densities
         of $^4$He and $^9$Be, respectively, calculated by the AMD.
         Here, the densities are shown by taking integrals over 
         the coordinate $z$: $\int dz \rho(x,y,z)$.}
\end{figure}

\begin{figure}[t]
\includegraphics[width=0.40\textwidth]{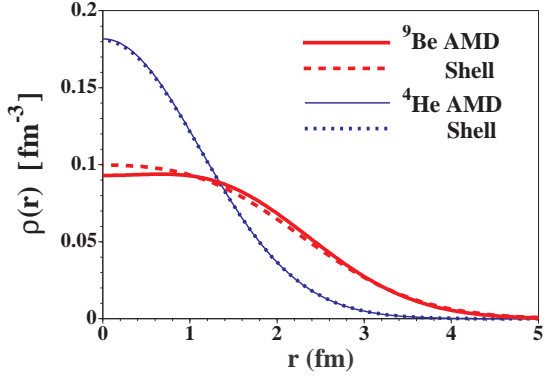}
\vspace{-0.2cm}
\caption{(Color online) Coordinate-space densities are shown for 
$^9$Be ($^4$He) in the AMD and shell models by the wide-solid (narrow-solid) 
and dashed (dotted) curves, respectively. The densities 
are integrated over the angles $\theta$ and $\phi$ for showing
the curves in this figure. The clustering structure 
in the AMD gives rise to a modification of the density distribution 
in $^9$Be, whereas both densities are the same in $^4$He.}
\label{fig:r-density}
\end{figure}

First, nuclear densities are shown in the AMD model.
The focused nucleus is $^9$Be for investigating the anomalous EMC 
effect in the structure function $F_2$, so that
coordinate-space densities are shown in Fig. \ref{fig:amd-4he-9be} 
for this $^9$Be nucleus as well as $^4$He, as an example, 
for comparison. 
It is interesting to find two density peaks within the $^9$Be nucleus, 
whereas the $^4$He density is a monotonic distribution.
As mentioned in Sec. \ref{formalism}, it is the advantage of
the AMD method that it does not assume any specific structure,
cluster- or shell-like configuration, on nuclei.
The $^4$He is a tightly bound nucleus and it is well
described by the usual shell-like structure, which is judged 
by the monotonic density distribution in Fig. \ref{fig:amd-4he-9be}.
However, the situation is apparently different in $^9$Be. 
The figure suggests that two dense regions exist
in $^9$Be although such a phenomenon does not exist in the shell model.
It indicates that the $^9$Be nucleus consists of two $\alpha$-like clusters
with surrounding neutron clouds. This clustering could produce 
different nuclear medium effects from the ones expected by the shell model.
In particular, it could influence the nucleon momentum distribution,
eventually quark momentum distributions, 
within the $^9$Be nucleus. Furthermore, dense regions could alter 
the internal structure of the nucleon.

Next, coordinate-space densities are compared in both AMD and shell models 
in Fig. \ref{fig:r-density} by taking averages over the polar and 
azimuthal angles $\theta$ and $\phi$. Although the $^4$He densities
are same in both models, they are different in $^9$Be.
Since the angular integrals have been
done, the cluster structure is no longer apparent in the AMD density
of $^9$Be in Fig. \ref{fig:r-density}.
However, the cluster effects are reflected in the slightly larger
densities at $r \sim$2 fm and the depletion at $r=0$,
due to the existence of two separate clusters.

\begin{figure}[t]
\includegraphics[width=0.40\textwidth]{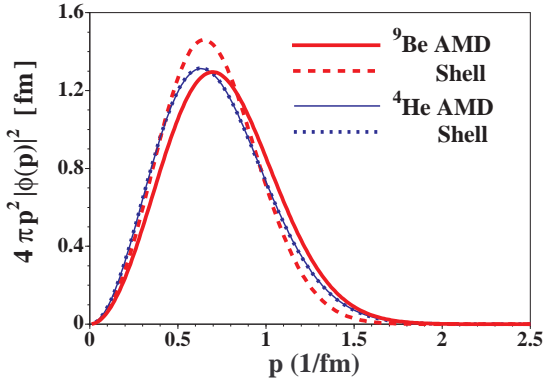}
\vspace{-0.2cm}
\caption{(Color online) Momentum-space densities are shown 
for $^4$He and $^9$Be in the AMD and shell models. The clustering
structure in $^9$Be gives rise to an excess 
of high-momentum components in the AMD.}
\label{fig:p-density}
\end{figure}

Instead of the coordinate-space density, the momentum-space density
$| \phi (\vec p_N) |^2$ is used for calculating the light-cone
momentum distribution by Eq. (\ref{eqn:f(y)}). Calculated
momentum-space densities are shown in Fig. \ref{fig:p-density}
for the nuclei $^4$He and $^9$Be. We explained in Sec. \ref{data}
that the same radii are taken in both AMD and shell models. 
As a result, both momentum distributions of $^4$He are almost
the same. However, the distributions
are much different in $^9$Be. {\it It is important to find that 
the momentum distribution of the AMD is shifted toward 
the high-momentum region in $^9$Be because of 
the clustering structure.} This is caused by the fact that
the dense regions, namely the two clusters, are formed 
within the $^9$Be nucleus. If nucleons are confined
in the small space regions of the clusters, it leads to 
an increase of high momentum components,
which is clearly shown in Fig. \ref{fig:p-density}.

\begin{figure}[t]
\includegraphics[width=0.40\textwidth]{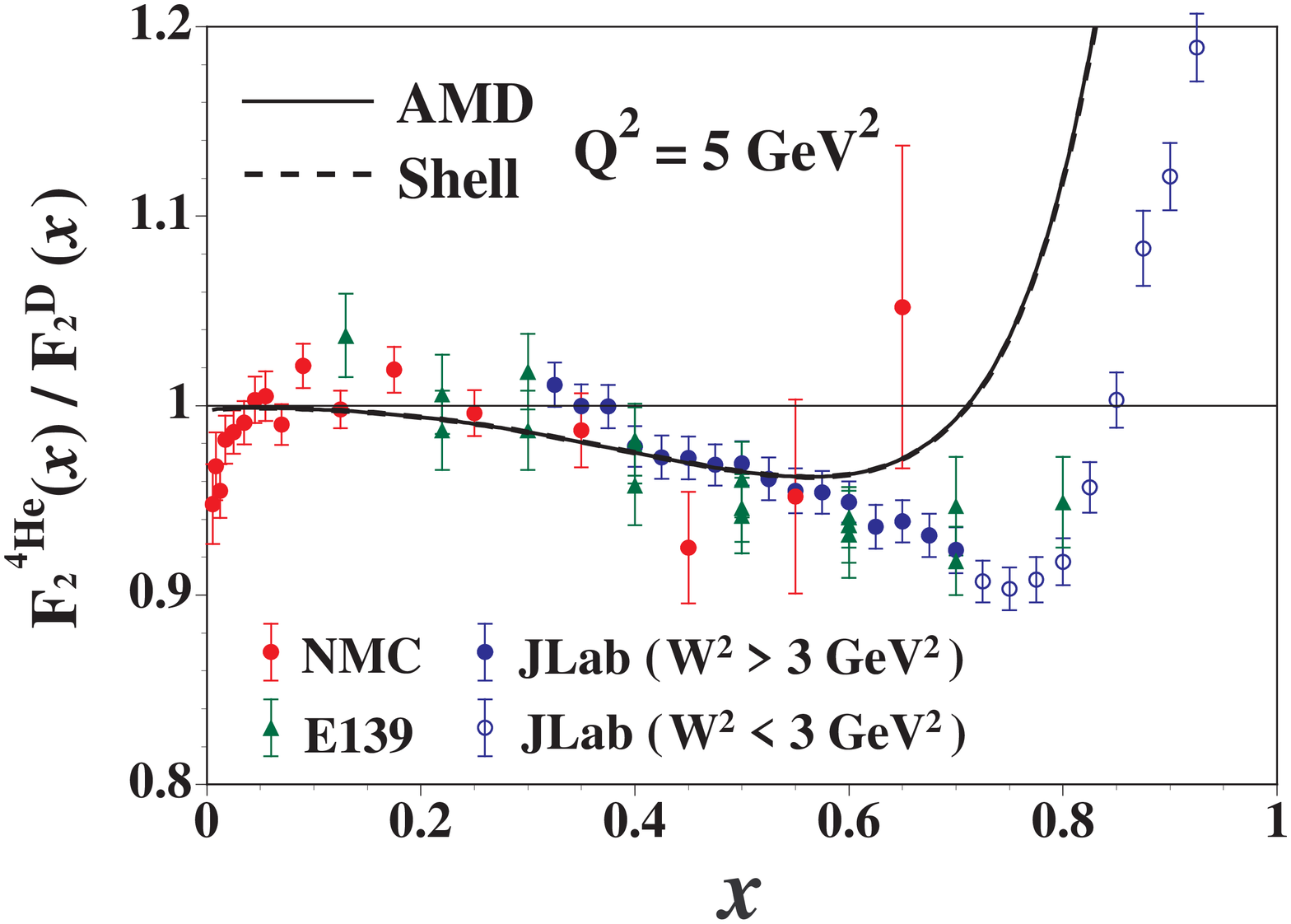}
        \vspace{-0.25cm}
\caption{\label{fig:4he}
(Color online) Theoretical structure-function ratios 
   $F_2^{^4 He}/F_2^D$ are compared with experimental data 
   of SLAC-E139 \cite{slac94-e139}, NMC \cite{nmc95}, 
   and JLab \cite{jlab-2009}. The solid and dashed curves
   indicate AMD and shell model results, respectively, calculated
   at $Q^2$=5 GeV$^2$; however, both curves overlap each other.
   The experimental data are taken at various $Q^2$ points.}
\vspace{0.5cm}
\includegraphics[width=0.40\textwidth]{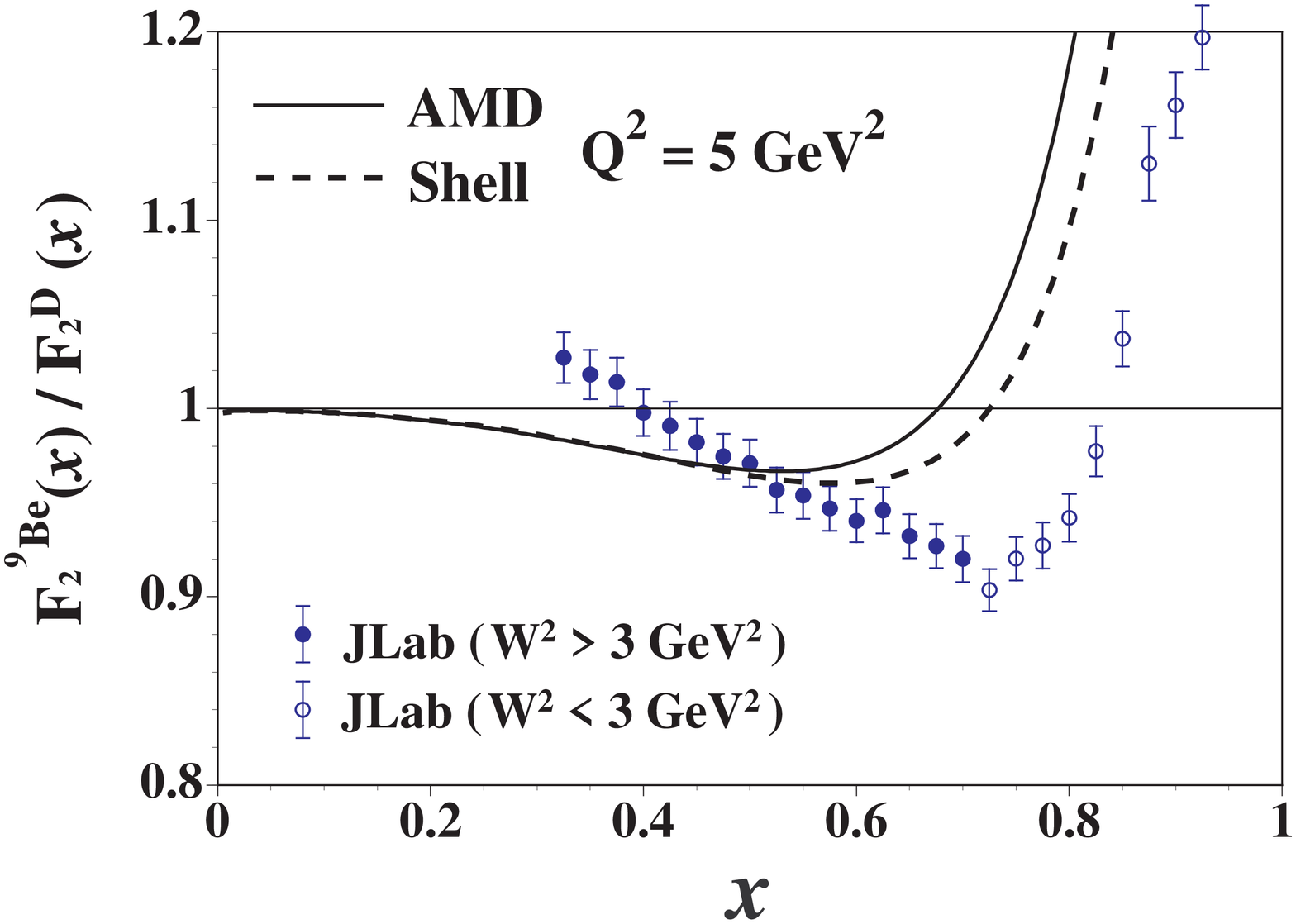}
        \vspace{-0.25cm}
\caption{\label{fig:9be}
(Color online) Comparison with JLab experimental data of $F_2^{^9 Be}/F_2^D$.
Notations are the same as the ones in Fig. \ref{fig:4he}. The differences
between the solid (AMD) and dashed (shell model) curves are now clear
in $^9$Be. Isoscalar corrections are applied as explained 
in the main text.}
\end{figure}

Now, using the obtained momentum distributions together 
with Eqs. (\ref{eqn:w-convolution}), (\ref{eqn:f(y)}), 
and (\ref{eqn:sp}), we calculate the nuclear structure
functions. The structure function of $^9$Be is shown together
with the one of $^4$He as an example of non-cluster-like nuclei
in order to illustrate clustering effects on the structure
function of $^9$Be.
In Figs. \ref{fig:4he} and \ref{fig:9be}, our theoretical ratios 
$F_2^{^4 He}/F_2^D$ and $F_2^{^9 Be}/F_2^D$ are compared with 
the available experimental data of the
SLAC (Stanford Linear Accelerator Center)-E139 \cite{slac94-e139}, 
NMC (New Muon Collaboration) \cite{nmc95}, and JLab \cite{jlab-2009}.
The AMD and shell-model ratios are shown by the solid and dashed
curves, respectively, and they are calculated at a fixed $Q^2$
point ($Q^2=5$ GeV$^2$). Experimental data are taken at various $Q^2$
points, and only the data with $Q^2 \ge 1$ GeV$^2$ are shown
in Figs. \ref{fig:4he} and \ref{fig:9be}.
The JLab measurements include the data with small invariant mass $W$, where
the process is not considered to be deep inelastic. Therefore,
the data with $W^2 <3$ GeV$^2$ are shown by the open circles.
In showing the ratios of non-isoscalar ($Z \ne N$) nuclei,
isoscalar corrections are applied in Ref. \cite{jlab-2009}
by including smearing corrections. Since we cannot access 
to the specific smearing corrections in the JLab analysis,
we simply used the isoscalar corrections 
$[(F_2^p+F_2^n)/2]/[ (Z F_2^p + N F_2^n)/A]$, where $F_2^p$ and
$F_2^n$ are the structure functions of the proton and neutron,
respectively, by using the PDFs of the MSTW08 \cite{mstw08} in 
the leading-order (LO) of $\alpha_s$. We have checked
that our corrections are almost the same as the corrections
in the JLab analysis in $^9$Be \cite{jlab-2009}.

From Figs. \ref{fig:4he} and \ref{fig:9be}, we find that our 
theoretical ratios have a tendency consistent with the data
in the sense that the ratio decreases at medium $x$ 
and it increases at large $x$. These decrease and increase
are caused by the nuclear binding and the nucleon's Fermi motion,
respectively, in our convolution picture.
However, it is also clear that the simple convolution description
is not sufficient to explain the whole experimental nuclear 
modifications because there are differences between the theoretical
curves and the data.

There are two major reasons for the differences. 
First, short-range nucleon-nucleon correlations
have not been included in calculating the spectral function
\cite{cfks-2007,cl-1990}. They change the theoretical ratios toward
the experimental data at $x=0.6-0.8$. 
The purpose of our studies is to investigate whether or not
a possible clustering signature appears in deep inelastic
lepton-nucleus scattering. Since this is the first attempt 
to investigate the cluster effects, we did not include 
such an effect. In future, we may consider to study more details.

Second, there could be a modification of nucleon itself
inside a nuclear medium. As explained in Refs. \cite{sumemc,roberts},
such a nucleon modification was originally proposed as a $Q^2$ rescaling
model. Nucleons could overlap in a nucleus since the average nucleon
separation and nucleon diameter are almost the same. 
The overlap then gives rise to a confinement radius change for quarks,
which appears as a modification of quark momentum distribution, 
namely a modification of the structure function $F_2$. 
A possible internal nucleon modification was investigated
in Refs. \cite{st1994,kp2010} in comparison with the data.  
Since it is not the purpose of this work to step into such details,
especially in comparison with the data, we leave it for our possible
future studies.

\begin{figure}[b]
\includegraphics[width=0.42\textwidth]{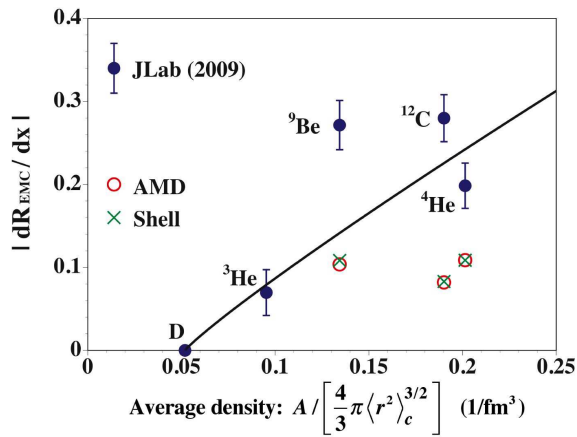}
\vspace{-0.2cm}
\caption{(Color online) Comparison with JLab data on
    the slope $| dR_{EMC} /dx |$, where $R_{EMC}=F_2^A/F_2^D$.
    The open circles and crosses are the theoretical slopes 
    calculated by the AMD and shell models, respectively. 
    The JLab data are shown by the filled circles with errors. 
    The abscissa is the average density $\rho$ defined by 
    $A/[4 \pi <r^2>_c^{3/2}/3]$ with the charge r.m.s. radius 
    $\sqrt{<r^2>_c}$. The curve indicates a smooth function
    $| \, dR_{EMC}/dx \, | = a (\rho-\rho_{_D})^b$
    to fit the JLab experimental data except for $^9$Be.}
\label{fig:drdx-1}
\end{figure}

The anomalous data was reported for $^9$Be by the JLab experiment 
\cite{jlab-2009} by taking a slope of the ratio $F_2^A/F_2^D$ 
with respect to the Bjorken variable $x$ in the region $0.35<x<0.7$.
As shown in Fig. 4 of Ref. \cite{jlab-2009},
the magnitude of the $^9$Be slope is too large to be expected
from its average nuclear density in comparison with the ones 
of other light nuclei. We calculate corresponding theoretical slopes
by taking the derivatives $d(F_2^A/F_2^D)/dx$ ($\equiv dR_{EMC}/dx$)
at $x=0.35$. The JLab data are plotted by the average density calculated 
by a Greens Function Monte Carlo method \cite{density-mc}
with the multiplication factor $(A-1)/A$ for removing
the struck nucleon. 

Although this theoretical density estimate would be reliable,
we first show the slope by a purely experimental quantity
by defining an average density as 
$A/[4 \pi \left < r^2 \right > _c^{3/2}/3]$, 
where $\sqrt{<r^2>_c}$ is the experimental charge r.m.s. radius,
in Fig. \ref{fig:drdx-1} instead of the specific theoretical density.
The experimental charge radii are taken from 
Table \ref{table:energies-radii}.
The theoretical slopes in the AMD and shell models
are shown by the open circles ($\bigcirc$) and crosses ($\times$), 
respectively, for $^4$He, $^9$Be, and $^{12}$C. Since the experimental
separation energy is not available for $^3$He, the theoretical
slopes are not calculated for $^3$He.
The JLab data are shown by the filled
circles with errors. In order to illustrate how the $^9$Be slope
deviates from the other nuclear ones, a curve is given
in Fig. \ref{fig:drdx-1} by fitting the data without
the $^9$Be data in a simple functional form,
$ | \, dR_{EMC}/dx \, | = a (\rho-\rho_{_D})^b$ where
$\rho = A/[4 \pi <r^2>_c^{3/2}/3]$ and $\rho_{_D}$
is the density of the deuteron. 
The parameters $a$ and $b$ are determined and
we obtain $| \, dR_{EMC}/dx \, | = 1.35 (\rho-\rho_{_D})^{0.906}$,
which is the curve in Fig. \ref{fig:drdx-1}.
It is obvious that the $^9$Be slope is anomalous in the sense
that the data significantly deviates from the curve.

The magnitudes of the theoretical slopes are rather small 
in comparison with the data, and they are about half 
or less of the experimental ones in Fig. \ref{fig:drdx-1}.
This was already obvious from Figs. \ref{fig:4he}
and \ref{fig:9be} that the magnitudes of the theoretical slopes
are smaller than the experimental ones. As explained, the differences
could be caused by the short-range correlations and internal
nucleon modifications. An interesting result is that
the clustering effects are not apparent in the slope
by looking at both AMD and shell-model results for $^9$Be
although there are some differences in the momentum
distributions of Fig. \ref{fig:p-density}
and in the structure functions of Fig. \ref{fig:9be}.
In other nuclei, both theoretical slopes are almost identical.
This is understood in the following way. 
In the medium-$x$ region, the nuclear modifications can be described
mainly by the first two moments of the nucleon momentum distribution 
$f(y)$. These moments are expressed by the average separation and kinetic
energies, $<\varepsilon>$ and $<T>$ \cite{kulagin}, which are similar
in both models. It leads to the small differences between the AMD and
shell models in the slope $dR_{EMC}/dx$.

The small difference between the AMD and shell-model slopes
in $^9$Be suggests us to look for another reason
to explain the anomalous JLab data. As we noticed in 
Fig. \ref{fig:amd-4he-9be}, the high density regions are created
locally in $^9$Be according to the AMD model. The higher
densities could contribute to extra nuclear modifications
in the structure function $F_2^A$ by an additional mechanism
which is not considered in our simple convolution picture.
In order to find such a possibility, we plot the same slope
by taking the maximum local density as the abscissa.
The maximum density $\rho_{max}$, of course, depends on
theoretical models to describe the nuclei.
The maximum positions are located at $r=0$ for 
$^4$He in both AMD and shell models and also for $^9$Be
in the shell model. However, they are at different points
($r \ne 0$) for $^9$Be in the AMD due to the cluster structure
and for $^{12}$C in both models due to a $p$-wave contribution
to the density.

\begin{figure}[b]
\includegraphics[width=0.42\textwidth]{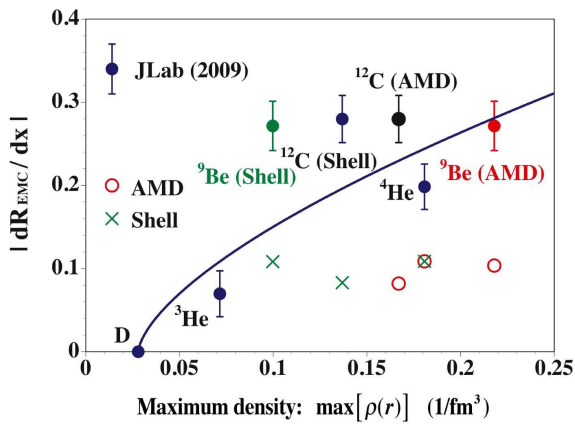}
\vspace{-0.2cm}
\caption{(Color online) Comparison with JLab data on $| dR_{EMC} /dx |$.
   The abscissa is the maximum local densities theoretically calculated 
   in the AMD and shell models. Two models produce different densities
   in $^9$Be and $^{12}$C, so that the JLab and theoretical slopes
   are plotted at different density points. The curve indicates
   a smooth function to fit the JLab experimental data except 
   for $^9$Be with the $^{12}$C data at the shell-model density.}
\label{fig:drdx-2}
\end{figure}

The slopes are shown in Fig. \ref{fig:drdx-2} by taking
the maximum local density $\rho_{max}$ as the abscissa. 
The maximum densities are almost the same in $^3$He and $^4$He,
so that they are plotted at the same position of $\rho_{max}$.
However, they differ in $^9$Be and $^{12}$C.
Although the difference between $\rho_{max}$ (AMD) and
$\rho_{max}$ (Shell) in $^{12}$C is not as large as 
the one in $^9$Be, it seems that there exist some 
clustering effects also in $^{12}$C.
The JLab data of $^9$Be and $^{12}$C and the theoretical slopes
are plotted at two different density points of the AMD and shell models.
The curve indicates a fit to the JLab data 
with the $^{12}$C data at the shell-model density point
by excluding the $^9$Be data. It is given by
$ | \, dR_{EMC}/dx \, | = 0.821 (\rho_{max}-\rho_{max {_D}})^{0.646}$,
where $\rho_{max {_D}}$ is the maximum density for the deuteron.

In the usual convolution calculation, the spectral function is given by
the averaged nuclear density distribution, and thus the inhomogeneity of 
the nuclear density is washed out. In such calculation, the average 
nuclear density of $^9$Be is lower than that of $^{12}$C or $^4$He 
as shown in Fig. \ref{fig:drdx-1}, which is the origin of the ``anomalous" 
EMC ratio of $^9$Be observed at the JLab. However, the maximum local 
density of $^9$Be is, as shown in Fig. \ref{fig:drdx-2}, higher than that 
the ones of $^4$He and $^{12}$C, and the EMC ratio of $^9$Be can be 
treated ``normally".  
Here, the ``normal" means that the result of $^9$Be is consistent with
the smooth curve determined by the EMC results of other nuclei, and thus 
$^9$Be does not have anomalous dependence on the nuclear density anymore. 
In this sense, the anomalous $^9$Be result is ``explained" as a normal
one by the maximum local density; however, it does not mean that physics 
mechanism is clarified.

What we emphasize here is that the EMC ratio or the nuclear 
structure function itself could consist of the mean conventional part 
and the remaining one depending on the maximum local density. The remaining
part is surely associated with the inhomogeneity of the nuclear density,
before taking the average of nuclear wave function, given by the nuclear 
cluster structure. It could be nuclear-medium modification of the nucleonic
structure function. It is well known that the cluster structure is 
well developed in Be and that the light mass region with $A<20$ is 
very suitable to study the cluster structure. Therefore, it is
reasonable that, although the cluster-structure effect in the EMC 
ratio is not seen in the medium and large nuclei so far, 
we can now observe it in the beryllium isotope region. 
Such cluster structure will be investigated in the light-mass region 
by future JLab experiments \cite{jlab-cluster-exp}.

It is interesting to find that {\it the ``anomalous" JLab data for $^9$Be 
can be explained if it is plotted by the maximum local density at 
the cluster positions} because the $^9$Be (AMD) data is very close to
the curve in Fig. \ref{fig:drdx-2}.
On the other hand, the $^9$Be data remains anonymous if it is plotted
by the shell-model density because the $^9$Be (Shell) data significantly
deviates from the curve. Such a tendency also exists in $^{12}$C but 
it is not as serious as the $^9$Be case. 
If the average nuclear density of $^9$Be is used in showing 
the slope data, the clustering effects are not clearly reflected.
Here, it is important to point out that the $^9$Be data agrees with 
the other nuclear data if they are plotted as a function of 
the maximum density. This fact implies that
{\it the physics mechanism associated with the high densities,
for example due to the clusters in $^9$Be,
could be the origin for explaining the nuclear-modification slopes 
of the $F_2^A$ structure functions.}
One of the possible mechanisms is the modification of internal 
nucleon structure caused by nuclear medium effects 
at the high density regions.

\section{Summary}\label{summary}

Nuclear modifications of structure function $F_2$ were investigated
for finding a possible signature of clustering structure in nuclei. 
The convolution model was used for describing nuclear structure functions,
where momentum distributions of the nucleon were calculated in
the AMD and shell models. According to the AMD, the $^9$Be nucleus
has a clear clustering structure of two $\alpha$-like clusters with
a surrounding neutron. Because of the cluster formation in $^9$Be,
high-momentum components increase in the nuclear wave function of
the AMD in comparison with the distribution of the shell-model one.
Although there are some differences between the structure functions of 
$^9$Be in the AMD and shell models, the differences are rather small 
in our simple convolution picture.
Therefore, an anomalous EMC effect found for $^9$Be at JLab should
come from other effects such as the internal nucleon modification 
due to the high-density regions created by the clustering.

The following points are the major results in this work:
\begin{itemize}
\vspace{-0.15cm}
\item[(1)] For the first time, the nuclear structure functions $F_2^A$
  are calculated in a model with clustering structure in nuclei.
  Then, they are compared with the structure functions of the shell model
  to clarify the clustering effects.
\vspace{-0.15cm}
\item[(2)] The clustering configuration in the $^9$Be nucleus produces 
  high-momentum components in the nuclear wave function. It leads 
  to a modification of the light-cone momentum distribution 
  for nucleons in the $^9$Be nucleus.
\vspace{+0.12cm}
\item[(3)] Because of the high-momentum components due to the cluster 
  formation, the nuclear structure functions $F_2^A$ are modified; however,
  the modifications are not very large within the simple convolution
  description.
\vspace{-0.15cm}
\item[(4)] The anomalously large nuclear effect for the slope
  $| \, d(F_2^A/F_2^D)/dx \, |$ of the $^9$Be nucleus observed at JLab
  can be explained if the slope is plotted by 
  the maximum local density calculated in the theoretical model (AMD)
  with clustering structure.
\vspace{-0.15cm}
\item[(5)] Since the nuclear-modification slopes are explained 
  by the maximum densities of nuclei, the physics mechanism of
  the anomalous nuclear effect could be associated with the high
  densities in the clusters of $^9$Be. 
  This fact implies that internal nucleon modifications 
  due to the high densities could be the origin of the $^9$Be anomaly,
  although careful estimations should be made on effects 
  of short-range nucleon-nucleon correlations.  
\end{itemize}
\vspace{-0.15cm}

This work is the first attempt to connect the DIS structure functions
to the clustering structure in nuclei.
This kind of research field is an unexplored area, and further 
theoretical studies are needed for clarifying clustering effects
in the structure functions.

\begin{acknowledgements}
\vspace{-0.3cm}
The authors thank J. Arrington, C. Ciofi degli Atti, A. Daniel, 
A. Dote, D. Gaskell, N. Itagaki, H. Morita, T. Noro, N. Shimizu, 
and P. Solvignon for communications and suggestions.
\end{acknowledgements}

\vspace{-0.3cm}



\begin{thebibliography}{00}
\bibitem{emc03} J. J. Aubert {\it et al.} (European Muon Collaboration),
                        Phys. Lett. B {\bf 123}, 275 (1983).
\bibitem{npdfs} M. Hirai, S. Kumano, and M. Miyama,
                       Phys. Rev. D {\bf 64}, 034003 (2001); 
                M. Hirai, S. Kumano, and T.-H. Nagai,
                       Phys. Rev. C {\bf 70}, 044905 (2004); 
                                    {\bf 76}, 065207 (2007). 
\bibitem{recent-npdfs} K. J. Eskola, H. Paukkunen, and C. A. Salgado,
                JHEP {\bf 04}, 065  (2009) and references therein.
\bibitem{sumemc}  D. F. Geesaman, K. Saito, and A. W. Thomas,
                        Ann. Rev. Nucl. Part. Sci. {\bf 45}, 337 (1995).
\bibitem{uvdv-exp} Jefferson Lab PAC-34 proposal, PR12-09-004 (2008).
\bibitem{nutev-anomaly} G. P. Zeller {\it et al.} (NuTeV Collaboration),
                        Phys. Rev. Lett. {\bf 88}, 091802 (2002);
                        Erratum 
                        {\bf 90}, 239902 (2003). 
\bibitem{sinth} S. Kumano, Phys. Rev. D {\bf 66}, 111301 (2002); 
                M. Hirai, S. Kumano, and T.-H. Nagai,
                            Phys. Rev. D {\bf 71}, 113007 (2005);
                K. J. Eskola and H. Paukkunen, JHEP {\bf 0606}, 008 (2006);
                I. C. Clo\"et, W. Bentz, and A.W. Thomas,
                            Phys. Rev. Lett. {\bf 102}, 252301 (2009).
\bibitem{local-emc} S. Kumano and F. E. Close, 
                        Phys. Rev. C {\bf 41}, 1855 (1990);
                    C. Ciofi degli Atti and S. Liuti, 
                        Nucl. Phys. A {\bf 532}, 235 (1991);                         
                    C. Ciofi degli Atti, L. P. Kaptari, and S. Scopetta, 
                        Eur. Phys. J. A {\bf 5}, 191 (1999).
\bibitem{local-exp} C. Ciofi degli Atti, talk at the workshop on
                    the Jefferson Laboratory Upgrade to 12 GeV,
                    Seattle, USA, Oct. 27, 2009,
                    http://www.int.washington.edu/talks/WorkShops /int\_09\_3/;
                    K. Hafidi {\it et al.}, 
                        Jefferson Lab PAC-35, Letter of Intent (2009).
\bibitem{neutrino-A} I. Schienbein {\it et al.},
                        Phys. Rev. D {\bf 77}, 054013 (2008);
                     M. Hirai, S. Kumano, and K. Saito, 
                        AIP Conf. Proc. {\bf 1189}, 269 (2009);                      
                     H. Paukkunen and C. A. Salgado,
                        JHEP {\bf 1007}, 032 (2010).
\bibitem{mstw08} A. D. Martin, W. J. Stirling, R. S. Thorne,
                 and G. Watt,  Eur. Phys. J. C {\bf 63}, 189 (2009).
                        The LO PDFs are used in this work.
\bibitem{minerva} For the MINER$\nu$A project, 
                           see http://minerva.fnal.gov/.
\bibitem{e906} E906 experiment      
                at http://p25ext.lanl.gov/e866/e866.html.
\bibitem{j-parc} See http://j-parc.jp/index-e.html for the J-PARC project.
                 S. Kumano, Nucl. Phys. A {\bf 782}, 442 (2007); 
                 AIP Conf. Proc. {\bf 1056}, 444 (2008).
\bibitem{jlab-2009} J. Seely {\it et al.}, 
	                    Phys. Rev. Lett. {\bf 103}, 202301 (2009).
\bibitem{amd-1995} Y. Kanada-En'yo, H. Horiuchi, and A. Ono,
                        Phys. Rev. C {\bf 52}, 628 (1995).
\bibitem{amd-1997} A. Dote, H. Horiuchi, and Y. Kanada-En'yo,
                        Phys. Rev. C {\bf 56}, 1844 (1997).
\bibitem{sf-introduction} R. Devenish and A. Cooper-Sarkar,
                 {\it Deep Inelastic Scattering}
                 (Oxford University Press, 2004), pp. 57-60 \& 370.
\bibitem{roberts} R. G. Roberts, 
                 {\it The Structure of the Nucleon}
                 (Cambridge University Press, 1993), pp. 8-12 \& 144-153. 
\bibitem{sk-pr} S. Kumano, Phys. Rep. {\bf 303}, 183 (1998).
\bibitem{ek03}  M. Ericson and S. Kumano, 
                        Phys. Rev. C {\bf 67}, 022201 (2003). 
\bibitem{kk08} T.-Y. Kimura and S. Kumano,
                        Phys. Rev. D {\bf 78}, 117505 (2008). 
\bibitem{li-liu-brown-88} G. L. Li, K. F. Liu, and G. E. Brown,
                        Phys. Lett. B {\bf 213}, 531 (1988).
\bibitem{cfks-2007}
      C. Ciofi degli Atti, L. L. Frankfurt, L. P. Kaptari, and M. I. Strikman, 
                  Phys. Rev. C {\bf 76}, 055206 (2007).
\bibitem{kulagin} S. V. Akulinichev, S. A. Kulagin, and G. M. Vagradov,
                    Phys. Lett. B {\bf 158}, 485 (1985);
                  S. A. Kulagin, Nucl. Phys. A {\bf 500}, 653 (1989);
                  S. A. Kulagin and R. Petti, 
                       Nucl. Phys. A {\bf 765}, 126 (2006). 
\bibitem{cl-1990} C. Ciofi degli Atti and S. Liuti, 
                        Phys. Rev. C {\bf 41}, 1100 (1990).
                  See also C. Ciofi degli Atti and S. Simula,
                        Phys. Rev. C {\bf 53}, 1689 (1996).
\bibitem{mt-1991} H. Morita and T. Suzuki, 
                        Prog. Theor. Phys. {\bf 86}, 671 (1991). 
\bibitem{amd-intro} Y. Kanada-En'yo, M. Kimura, and H. Horiuchi,
                        C. R. Physique {\bf 4}, 497 (2003);
                    M. Kimura, A. Dote, A. Ohnishi, and H. Matsumiya,
                        Genshikaku Kenkyu {\bf 53}, Supplement 2, 50 (2009) 
                        (in Japanese).
\bibitem{fmd-intro} H. Feldmeier and J. Schnack, 
                        Rev. Mod. Phys. {\bf 72}, 655 (2000).                         
\bibitem{mc-8be-2000} R. B. Wiringa, S. C. Pieper, J. Carlson,
        and V. R. Pandharipande, Phys. Rev. C {\bf 62}, 014001 (2000). 
\bibitem{3odd} R. Tamagaki, Prog. Theo. Phys. {\bf 39}, 91 (1968).
\bibitem{ono-1992} A. Ono, H. Horiuchi, T. Maruyama, and A. Ohnishi,
                        Prog. Theo. Phys. {\bf 87}, 1185 (1992).
\bibitem{rs-book} P. Ring and P. Schuck, {\it The Nuclear Many-Body Problem}
                    (Springer-Verlag, 2004).
\bibitem{nushell} See http://knollhouse.org/.
\bibitem{fw-book} A. L. Fetter and J. D. Walecka,
         {\it Quantum Theory of Mary-Particle Systems}
         (McGraw-Hill, 1971), pp. 508-511.
\bibitem{deuteron} See pp.66-69 in R. Machleidt, K. Holinde, and C. Elster, 
                        Phys. Rept. {\bf 149}, 1 (1987). 
\bibitem{audia-2003} G. Audi, A. H. Wapstra, and C. Thibault, 
                        Nucl. Phys. A {\bf 729}, 337 (2003).                         
\bibitem{rms-d} R. C. Barrett and D. F. Jackson
                  pp.146-147 in {\it Nuclear sizes and structure} 
                  (Oxford: Clarendon Press , 1977).
\bibitem{rms-3he} A. Amroun {\it et al.},  
                        Nucl. Phys. A {\bf 579}, 596 (1994);
                  J. Golak {\it et al.}, Phys. Rep. {\bf 415}, 89 (2005).
\bibitem{rms-4he} R. Roth {\it et al.}, Nucl. Phys. A {\bf 745}, 3 (2004).
\bibitem{rms-9be} F. Ajzenberg-Selove, Nucl. Phys. A {\bf 490}, 1 (1988).
\bibitem{rms-12c} H. De Vries, C. W. De Jager, and C. De Vries,
                     Atom. Data Nucl. Data Tabl. {\bf 36}, 495 (1987).                         
\bibitem{tyren-1966} H. Tyr\'en {\it et al.}, Nucl. Phys. {\bf 79}, 321 (1966).
\bibitem{separation} S. Frullani and J. Mougey, 
                        Adv. Nucl. Phys. {\bf 14} (1984) 1. See page 194.
\bibitem{us-1988} T. Uchiyama and K. Saito, 
                        Phys. Rev. C {\bf 38}, 2245 (1988).
\bibitem{3he4he} H. Morita and T. Noro, personal communications (2010)
                 on separation energies in $^3$He and $^4$He.
\bibitem{slac94-e139} J. Gomez {\it et al.},    
                         Phys. Rev. D {\bf 49}, 4348 (1994).
\bibitem{nmc95} P. Amaudruz {\it et al.},       
                    Nucl. Phys. {\bf B441}, 3 (1995);
                M. Arneodo {\it et al.},
                    {\it ibid.} {\bf B441}, 12 (1995).
\bibitem{st1994} K. Saito and A.W. Thomas, 
                       Nucl. Phys. A {\bf 574}, 659 (1994); 
                 I. C. Clo\"et, W. Bentz, and A. W. Thomas, 
                       Phys. Lett. B {\bf 642}, 210 (2006).
\bibitem{kp2010} S. A. Kulagin and R. Petti, arXiv:1004.3062 [hep-ph].
\bibitem{density-mc} S. C. Pieper and R. B. Wiringa, 
                       Annu. Rev. Nucl. Part. Sci. {\bf 51}, 53 (2001).
\bibitem{jlab-cluster-exp} Jefferson Lab PAC-35 proposal, PR12-10-008 (2009).

\end{thebibliography}
\end{document}